# Absolute light yield measurement of NaI:Tl crystals for dark matter search

N.T. Luan[1], H.J. Kim[1*], H.S. Lee[2], J. Jegal[3], L.T. Truc[1], A. Khan[4], N.D. Ton[1]

[1] Department of Physics, Kyungpook National University, Daegu, Korea

[2] Center for Underground Physics, Institute for Basic Science (IBS), Daejeon, Korea

[3] Department of Radiation Oncology, Seoul National University Hospital, Seoul, Korea

[4] Department of Physics, Faculty of Arts and Sciences, Najran University, Najran, Saudi Arabia

**Abstract** The NaI:Tl crystals were early investigated and used for wide application fields due to high light yield and crystal growth advantages. So far, the absolute light yields of NaI:Tl crystal have typically been known to be 40 ph/keV. However, it varies widely, far from the theoretical estimation. Since the high light yield and better sensitivity of NaI:Tl crystal is important for low mass dark matter search. Therefore, it is necessary to use high light NaI:Tl crystal, and absolute light yield should be measured with accuracy. In this work, we use the single photoelectron technique for measuring the absolute light yield of 35 NaI:Tl crystals with various sizes from different vendors. There are several high-quality crystals from the COSINE-100 experiment and commercial companies in these crystals. The theoretical estimation and GEANT4 optical simulation have been studied to investigate the PMT optics. Results show the essential role of this correction in avoiding overrated light yield values. The SPE technique using different PMT was compared to the photodiode and avalanche photodiode methods. A 10% systematic error was obtained. Our results show the excellent absolute light yield of NaI:Tl, at 59.4±5.9 ph/keV, while the theoretical predicted light yield is around 70 ph/keV. The evaluation with NaI:Tl crystals in the COSINE-100 experiment has been performed. The six crystals in the COSINE-100 experiment have a high light yield. Based on our results, the light loss of encapsulation needs to be improved, especially for the big-size crystals.

## 1 Introduction

Dark Matter, accounting for 84% of the universe's matter, is one of the deepest mysteries of our time. Its existence, first postulated based on astronomical observations, remains unconfirmed despite its ubiquity, thus deepening the mystery [1]. A multitude of theoretical explorations have given rise to several experimental strategies for dark matter detection. Based on the Weakly Interacting Massive Particles (WIMPs), the direct detection approach stands out as a promising candidate for dark matter searches [1].

Despite over three decades of searching for WIMP, definitive evidence for the existence of dark matter remains elusive. However, XENON-1T and DAMA/LIBRA experiments have reported intriguing results that warrant further scrutiny [2,3]. The XENON-1T experiment detected excess signals in 1300 kg of liquid Xenon, necessitating further verification to ascertain whether these signals are attributable to tritium contamination or WIMP interaction [2]. On the other hand, the DAMA/LIBRA experiment observed annual modulation using 250 kg NaI:Tl scintillation crystals, but the result could not be accounted for by known backgrounds [3]. Subsequent searches employing CsI:Tl scintillation crystals failed to corroborate the DAMA/LIBRA findings, sparking a global effort to replicate the DAMA/LIBRA observation using NaI:Tl detectors [4]. However, the DM-Ice 17 and NAIAD experiments did not detect annual modulation or WIMP signals [5,6]. The ongoing COSINE-100 and ANAIS-112 experiments impose stringent constraints on the DAMA/LIBRA results [7-10].

* e-mail: hongjoo@knu.ac.kr (corresponding author)

However, it is crucial to compare the background rates of these experiments with those of the DAMA/LIBRA experiments [11]. DAMA/LIBRA reported a background rate of approximately one count/kg/day/keV in the 1-6 keV range [3,11]. In contrast, the DM-Ice 17 and ANAIS experiments have a higher background rate. Specifically, the ANAIS-112 experiment reports a background rate around 3.2 times higher than DAMA/LIBRA, while the COSINE-100 experiment's background rate is 2.7 times higher [11]. This discrepancy suggests that these experiments need more effort to reproduce the DAMA/LIBRA result. New projects in the research and development (R&D) phase are exploring ways to reduce background noise and increase NaI:Tl mass to enhance detection sensitivity. For example, the COSINE-200 project plans to utilize 200 kg NaI:Tl crystals with an anticipated background rate of less than 0.5 counts/kg/day/keV [12]. Besides, SABER, PICOLON, and COSINUS projects are in R&D progress, aiming to reduce background levels for compatibility with DAMA/LIBRA [12-15].

Investigations into the background of experiments have underscored the pivotal role of NaI:Tl scintillation crystal quality. Achieving superior background levels necessitates high light yield and ultra-radiopure NaI:Tl crystals. Also, a higher light yield of NaI:Tl crystal could reduce the energy threshold and achieve higher sensitivity for dark matter searches and low mass dark matter searches. Each experiment conducts its own NaI:Tl research and development. Ultra-radiopure NaI:Tl crystals are characterized by a low internal background induced by radioactive isotopes, which can be regulated through content determination. The detector light output has been quantified in terms of photoelectrons per keV (pe/keV) rather than the absolute light yield of the crystal, which is measured in photons per keV (ph/keV). This implies that the true light yield nature of NaI:Tl crystals remain undetermined, potentially hindering the growth of NaI:Tl crystals and the enhancement of engineering techniques for optimal detector total light output. The absolute light yield of NaI:Tl crystal is heavily reliant on growth quality, which ranges from 20 ph/keV to 50 ph/keV and typical values around 40 ph/keV [16-20]. However, this falls short of the ultimate theoretical light yield prediction of 70 ph/keV [21]. Given the variety of techniques for absolute light yield measurement, each with its own strengths and weaknesses, there is a compelling motivation to explore suitable absolute light yield measurements for crystal quality control in research and development for WIMP searches.

In the pursuit of dark matter searches, NaI:Tl crystals are paired with photomultiplier tubes (PMTs). Consequently, measuring the absolute light yield of NaI:Tl crystals using PMTs is more effective than other photon detectors, such as silicon photodiode (Si-PD) or avalanche silicon photodiode (Si-APD). When utilizing PMTs, the absolute light yield can be measured by comparing the photopeak to the single photoelectron response, the so-called single photoelectron (SPE) technique. The SPE technique offers convenience and is less sensitive to temperature fluctuations than Si-PD or Si-APD techniques. It also excels in measuring larger-sized crystals. However, the optical reflectivity of PMTs plays a significant role in determining the absolute light yield as it pertains to the photocathode, influencing the actual quantum efficiency of PMT, which could potentially overestimate the absolute light yield. To comprehend this effect, this study examined PMT optical reflectivity through theoretical and simulation approaches. After applying the necessary corrections, we measured the absolute light yield of high-quality samples from COSINE-100 NaI:Tl. We tested several samples from Alpha Spectra Inc., along with other commercial or related samples, using various techniques to ascertain the reliability of the absolute light yield measurement using the SPE technique. Our measurements revealed excellent absolute light yields of NaI:Tl crystals. Furthermore, this study employed the same PMT as used in COSINE-100 modules. We estimated the quality of COSINE-100 NaI:Tl crystals and discussed the light loss.

## 2 Experimental setup

In this study, we measured 35 samples of NaI:Tl crystals, the details of which are provided in Table 1. The KNU+COSINE samples were grown using the Bridgman technique at Kyungpook National University, with ultra-radiopure NaI powder supplied by the COSINE-100 collaboration [22]. The COSINE-100 collaboration also contributed bare ASWSII-2, CUP KY01-2007, and CUP 036 from the Center for Underground Physics (CUP) samples grown in different conditions [8,23]. To explore the distribution of absolute light yield within a large crystal grown from COSINE-100 NaI:Tl R&D, we cut eleven samples of CUP KY01-2007 from various positions of a large crystal piece. We obtained a bare sample from an encapsulated one and procured an additional new sample from EPIC Crystal Co. [24]. For the bare samples from Alpha Spectra Inc. [25], one sample was extracted from the ingot, and four samples were prepared well by Alpha Spectra. We ordered two 3" encapsulated crystals from Alpha Spectra [25]. The BICRON [26] sample was derived from the NaI:Tl compact scintillator. Samples of the same type but different shapes can be seen in Fig. 1. The ASWSII-2 sample was encapsulated, as illustrated in Fig. 2, to enable measurements with Si-PD and Si-APD and to study the light loss due to encapsulation. The encapsulation process was carried out inside a metal glovebox filled with Argon to prevent the degradation of the crystal surface due to humidity. The crystal, which had been well-polished, was cleaned on the surface and wrapped with several layers of high-quality Teflon. The exposed surface of the wrapped crystal was swiftly attached to a clean quartz window using the EJ-500 optical cement [27]. The assembly was placed into a clean aluminum case and sealed with the same EJ-500 optical cement. The thickness of the Teflon was maximized to minimize the gap between the wrapped crystal and the aluminum wall. The remaining gap was filled by EJ-500 optical cement.

**Table 1** The measured NaI:Tl crystals

| Scintillation crystals | Dimension (mm³) | Number of samples |
|---|---|---|
| KNU+COSINE R&D | φ8x4-10x10x10 | 11 |
| Bare ASWSII-2 | 8x8x8 | 1 |
| Encapsulated ASWSII-2 | 8x8x8 | 1 |
| Bare CUP KY01-2007 | 8x8x8 | 11 |
| Bare CUP 036 | 8x8x8 | 1 |
| Bare Alpha Spectra 1 | 10x10x10 | 1 |
| Bare Alpha Spectra 2 | 10x10x10 | 4 |
| Bare EPIC | φ50.8x50.8 | 1 |
| Bare BICRON | φ76.2x76.2 | 1 |
| Encapsulated Alpha Spectra | φ76.2x76.2 | 2 |
| Encapsulated EPIC | φ50.8x50.8 | 1 |

We also evaluated additional air-stable crystal samples that exhibited varying luminescence emissions to validate the SPE technique like BGO, LYSO:Ce, GAGG:Ce, CsI:Tl, intrinsic $Cs_3Cu_2I_5$ and $Cs_3Cu_2I_5$:Tl crystals. In which the BGO, LYSO:Ce, and GAGG:Ce are commercial samples. CsI:Tl crystals were provided by the KIMS experiment [28]. The intrinsic and Tl-doped $Cs_3Cu_2I_5$ crystals were grown in Kyungpook National University using the Bridgman technique. The main peak emissions of these samples are provided in Table 2. The luminescence emission data was derived from X-ray-induced luminescence, measured under identical conditions. These samples facilitated the measurement of absolute light yield using Si-PD and Si-APD. Consequently, the sizes of these samples were prepared in the appropriate size to fit with the effective area of the Si-PD and Si-APD detector.

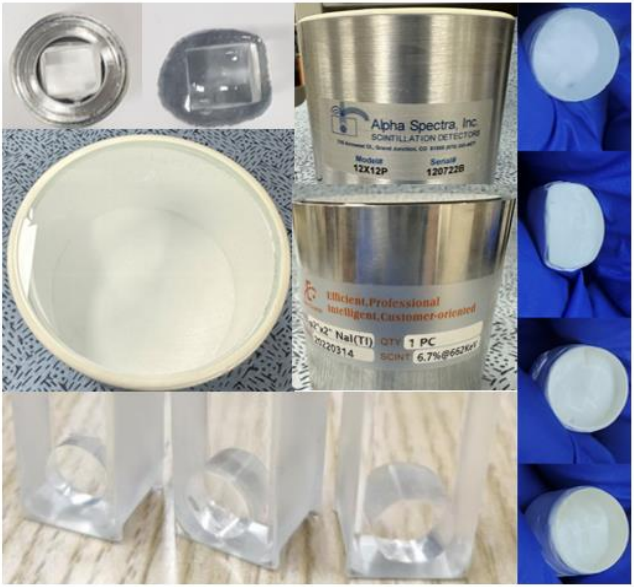

**Fig. 1.** NaI:Tl samples with different sizes and shapes.

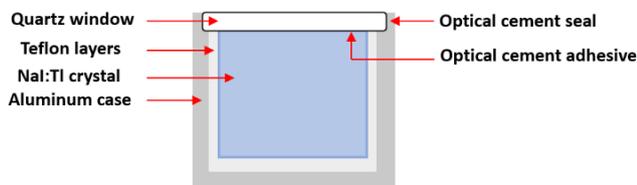

**Fig. 2.** ASWSII-2 NaI:Tl encapsulation

**Table 2** The measured crystals and their emission peak

| Scintillation crystals | Emission (nm) |
|---|---|
| BGO | 480 |
| LYSO:Ce | 410 |
| GAGG:Ce | 550 |
| CsI:Tl | 550 |
| $Cs_3Cu_2I_5$:Tl | 480 |
| $Cs_3Cu_2I_5$ | 430 |
| Encapsulated ASWSII-2 NaI:Tl | 415 |

The bare samples underwent a meticulous polishing process and were subsequently measured in a metal glove box filled with Argon, maintaining a humidity level below 100 ppm. Throughout the polishing phase, the samples were consistently coated with mineral oil, which was later cleaned off to measure pulse height spectra under radiation excitation. Data acquisition (DAQ) was carried out using the Hamamatsu R12669SEL photomultiplier tube (CUP-IBS) [8], which was located inside the metal glove box. The quantum efficiency of this PMT, as shown in Fig. 3, was compared to the X-ray luminescence of NaI:Tl samples. The voltage supply cable (SHV) and the anode output signal cable (BNC) were linked to an external 400 MHz flash analog-to-digital converter (FADC) (Korea Notice NFADC400) [29], located outside the metal glove box. Data-taking was facilitated by ROOT-based programs [30]. The high stable voltage, provided by 556 ORTEC Power Supply [31], was optimized to prevent pulse saturation and distortion while ensuring sufficient gain for a robust pulse signal. The experiment used the γ-ray excitation from 662 keV γ-rays of $^{137}$Cs source. A 20 μs window was established for data taking. A detailed setup can be seen in Fig. 4. Data analysis was conducted in Python utilizing the uproot package [32]. The typical pulse height spectra are depicted in Fig. 5.

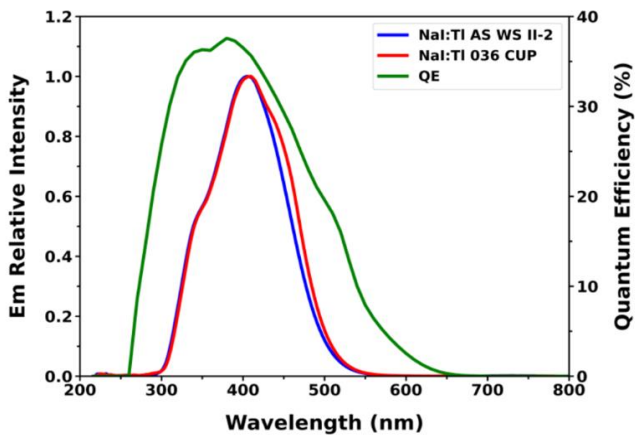

**Fig. 3**. X-ray induced emission of NaI:Tl crystals and the quantum efficiency of the PMT provided by the manufacturer.

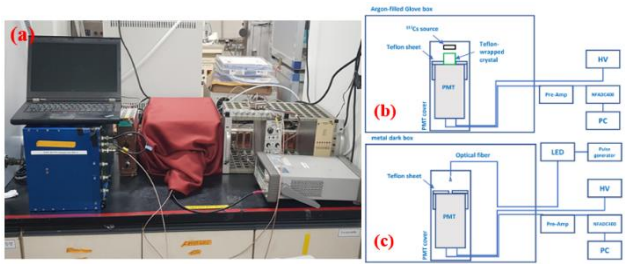

**Fig. 4.** The schematic diagrams of the experiment: (a) Our setup equipment, (b) pulse height spectrum measurement, and (c) SPE measurement.

The setup for the SPE measurement, as depicted in Fig. 4, involved directing a light-emitting diode (LED) source, controlled by a pulse generator, onto the surface of PMT. The LED pulse's frequency, duration, and applied voltage were carefully selected to ensure that the minimum number of photons reached the PMT. This was further verified by a high-performance spectrometer (QEpro, Optical Insight) [33] and Teledyne Lecroy wave runner 610Zi oscilloscope [34] with an excellent sampling rate with consideration of the distance between PMT surface and light source. As a result, the distribution of single photoelectrons was more pronounced than that of multiple photoelectrons. Given the low gain of the PMT, the pulse of a single photoelectron is relatively small compared to the pedestal noise. To address this, pre-amplification was applied tenfold. The multiplying factor was checked using BGO at different high voltages (HVs). The measured factors are 10.16±0.05, 10.25±0.07, and 10.24±0.08 for HVs of 1000 V, 1100 V, and 1200 V, respectively. These results indicate that the pre-amplification operated stably. Due to the pre-amplification, the clear photoelectron response was observed from the pedestal, suggesting an accurate determination of photoelectron response from the misidentified response from the pedestal level. The PMT was housed in a dark metal box and checked by the spectrometer, which showed a light level under the detection sensitivity of the spectrometer. The grounding and shielding of the setup were strongly considered to reduce the environmental electric noises. The threshold of FADC was set at the lowest level to record events. The time window of FADC was optimized in conjunction with the pulse duration and frequency to capture as many photons as possible. The SPEs were measured at HVs from 900V to 1200V. The LED was also switched off to measure dark current (DC) under the same conditions as the SPE measurement.

For the Si-PD and Si-APD techniques, we utilized specific devices under controlled conditions. The Hamamatsu S3590-08 photodiode was set to operate at 80V while the Hamamatsu S8664-55 avalanche photodiode was configured to run at 350 V [35]. We took into account the temperature dependence of photosensitivity during these measurements. The temperature was closely monitored and maintained at 17.0±0.5 ºC for the photodiode and 18.0±0.5 ºC for the avalanche photodiode. Furthermore, we utilized an Ortec pre-amplifier, notable for its electronic calibration constant of 2.1 ADC/fC, to enhance the precision of our measurement.

## 3 Integrated ADC and Single PE response

The photopeak of scintillation, excited by 661 keV γ rays from a $^{137}$Cs source, was determined via the pulse height spectrum. This spectrum was constructed based on the signal pulse from the PMT anode. The raw pulse underwent a process where the pedestal baseline was subtracted. This baseline was identified in the range preceding the rising edge of the anticipated signal, a region devoid of any signal. The mean and variance of the pedestal were extracted, enabling the rejection of any aberrant pedestal event based on multi-range pedestals. Tail pile-up pulses were detected using a first differential algorithm after subtracting the pedestal from the raw pulse. These pile-up pulses were subsequently discarded. This led to the derivation of pulse shape characteristics. The commencement position of the pulse was pinpointed as being 5% lower than the maximum amplitude of the pulse and higher than 3σ of the pedestal distribution. The time distribution between the starting and maximum amplitude positions across all pulses was utilized to mitigate head pile-up pulses. If the time of an individual pulse surpasses 5σ of this distribution, it is discarded. The selected pulses were calculated by integrating the ADC from the starting position. The constructed pulse height spectra of some samples are illustrated in Fig. 5.

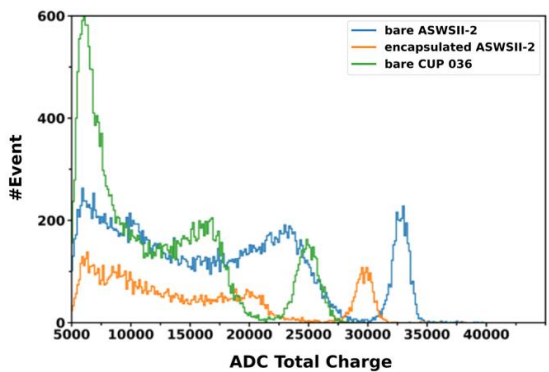

**Fig. 5.** The typical pulse height spectra.

For the measurement of the photoelectron, we utilized the same ADC window. Each recorded data set is expected to contain multiple photoelectron response signals, especially when the LED is activated. We employed a clustering method to search for potential response pulses from photoelectrons using clustering method. Clusters were identified, and the potential peak within each cluster was determined. Only clusters with a single peak were selected for further analysis. Due to pre-amplification, the baseline exhibits cosine behavior. The recognized clusters were subtracted from the data, and the remaining one was used to fit the baseline. A suitable baseline was fitted and subtracted, as shown in Fig. 6(a). This fitted baseline parameter was also used for the pedestal estimation of radio-excited signals of scintillation crystals using pre-amplification. The pedestal fluctuations, leading to misidentified peaks, were also identified and removed. The three general cases are illustrated in Fig. 6(b). The SPE charge was determined by considering the stability of the local pedestal, as shown in Fig. 6(c). Upon identifying the single peak, the integrated window was extended to both sides, and the charge of the corresponding pulse was calculated. The window-dependent charges are shown in Fig. 3(c). Based on this, the

SPE charge can be accurately calculated. In addition, unexpected poor pedestals or frequency noise could be detected and eliminated. The selected pulses were used to calculate the SPE charge. The number of recorded data sets was considered to achieve high statistics after selection. Our technique primarily focuses on the single photoelectron response.

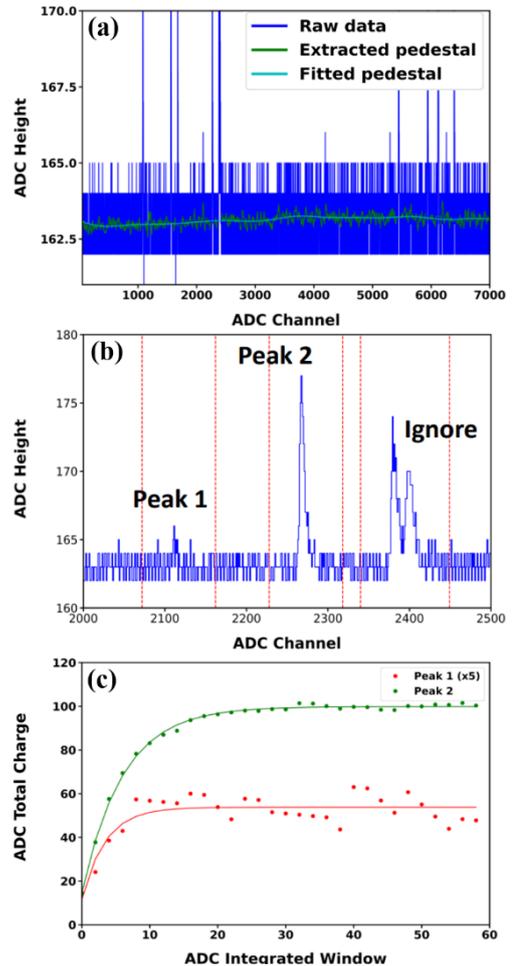

**Fig. 6.** The SPE determination: (a) pedestal baseline subtraction, (b) cluster finding, and (c) Integrated charge determination.

The single photoelectron (SPE) responses are illustrated in Fig. 7 regarding the charge and pulse height when the LED was turned on/off. Upon activation of the LED, an elevated photoelectron response is observed. The charge distributions of the SPE, corresponding to selected response pulses, are plotted in Fig. 8 across varying PMT operation voltages. At reduced voltages, the prominence of the pedestal is more noticeable, resulting in a less distinct valley.

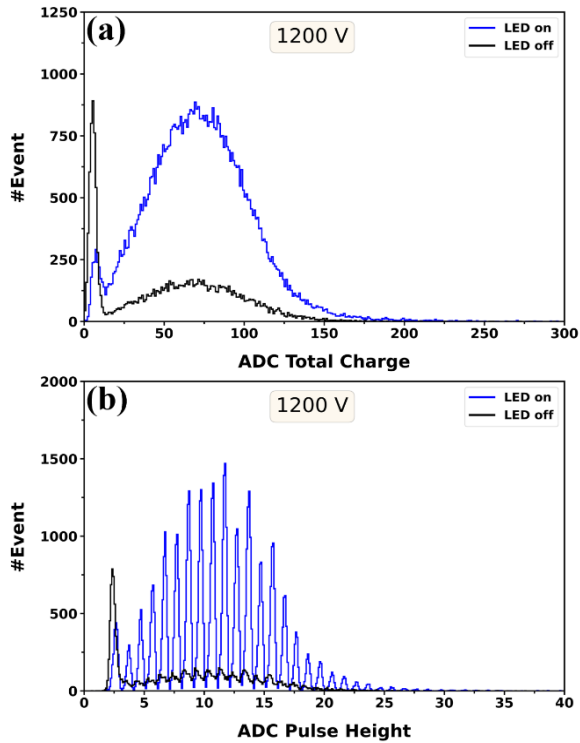

**Fig. 7.** The SPE (a) and height of photoelectron distribution (b) with/without LED using the Hamamatsu R12669-SEL (CUP-IBS) PMT.

In this study, we operated the Hamamatsu 3" R12669SEL PMT at a gain lower than $10^7$. It led to a challenge with detecting a single photoelectron at low gain due to the under-amplified photoelectrons, which are challenging to monitor [36]. However, this issue is not of significant concern for our specific objective of measuring absolute light yield. Considering the operation at low gain, the fitting function is formulated based on the photoelectron response model in Ref. [36]. It considers the ideal contribution of single photoelectrons (SPE), denoted as $f_{SPE}(x)$, which comprises Gaussian and exponential components. The exponential SPE distribution has a fraction of $P_E$ and is characterized by the slope ($x_e$). The mean ($x_0$) and standard deviation ($\sigma_0$) of SPE were introduced for the Gaussian SPE distribution. The distribution of SPE is considered concerning the pedestal as follows [37]:

$$f_{SPE}(x) = \frac{P_E}{x_E}e^{-\frac{x-x_P}{x_E}} + \frac{\sqrt{2}}{\sqrt{\pi}\sigma_0}\frac{1-P_E}{1+\mathrm{Erf}\left(\frac{x_0}{\sqrt{2}\sigma_0}\right)}e^{-\frac{(x-x_0-x_P)^2}{2\sigma_0^2}} \quad (1)$$

Here, $x_P$ presents the pedestal position. The term $\mathrm{Erf}\left(\frac{x_0}{\sqrt{2}\sigma_0}\right)$ is for the truncation of the Gaussian part. Additionally, the SPE distribution is convoluted with the pedestal white noise,

n(x), which can be explained by the Gaussian distribution with mean ($x_P$) and standard derivation ($\sigma_P$) as follows [37]:

$$n(x) = \frac{1}{\sqrt{2\pi}\sigma_P}e^{-\frac{(x-x_P)^2}{2\sigma_P^2}} \quad (2)$$

The multi-photoelectron distribution (NPE), $f_{NPE}(x)$, is also taken into account. It is considered with the Poisson distribution $P(n;\mu)$ where n is the number of photoelectron responses. The distribution could be explained by [37]:

$$f_{NPE}(x) = \sum_{n=2}^{N}\frac{P(n;\mu)}{\sqrt{2n\pi}\sigma_1}e^{-\frac{(x-nx_1-x_P)^2}{2n\sigma_1^2}} \quad (3)$$

In our case, there are small contributions of multiple photoelectrons, as suggested in Ref. [37], the approximation of $x_1$ and $\sigma_1$ could be written as follows:

$$x_1 \approx (1-P_E)x_0 + P_{EXP}x_E \quad (4)$$
$$\sigma_1 \approx (1-P_E)(\sigma_0^2+x_0^2) + 2P_Ex_e^2 - x_1^2 \quad (5)$$

The final form of photoelectron response distribution is:

$$f_{PE}(x) = P(0)\,n(x) + P(1)[f_{SPE}(x)\otimes n(x)] + f_{NPE}(x) \quad (6)$$

In which, $P(0)$ and $P(1)$ are the corresponding probability of detecting zero and one photoelectron, which follow the Poisson distribution. The fitting of photoelectron response distributions is illustrated in Fig. 8. The contribution of double photoelectron response is much less than that of single photoelectron response as we expected. The obtained value of $x_0$ was used for the absolute light yield determination. Additionally, the shape of SPE, as seen in Fig. 8, is characterized at low PMT gain [36].

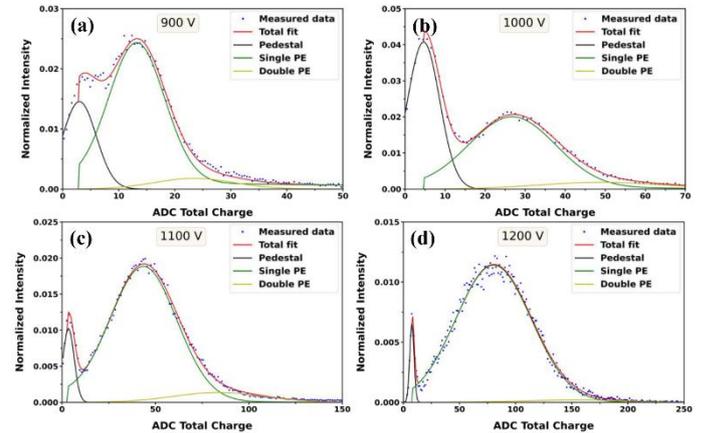

**Fig. 8.** The SPE peak fitting (a) HV of 900 V, (b) HV of 1000 V, (c) HV of 1100 V, and (d) HV of 1200 V.

## 4 Understanding of PMT optical processes

Visible photons, generated by the scintillation crystals under radio-excitation, traverse the PMT windows and induce the formation of photoelectrons within a thin layer of photocathode. This process, however, involves intricate mechanisms related to the multiple reflection and transmission of visible photons with the photocathode and other PMT construction materials. These interactions can influence the effective quantum efficiency, particularly in the case of scintillation with a reflection cover [19], which differs from the standard quantum efficiency measurement [38]. It's crucial to note that for the $4\pi$ emission of scintillation light, the reflection cover is indispensable for measuring the high light yield factor, which needs to be corrected for the absolute light yield measurement. Based on theoretical and experimental studies, the reflectance can be up to 20% or even higher [19,39-41]. Consequently, optical correction in PMT measurement is vital for accurate light yield measurement.

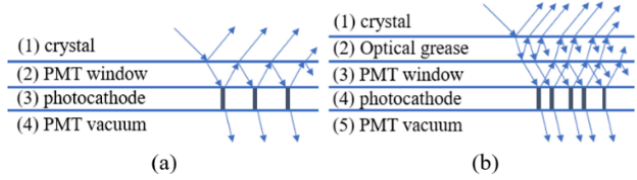

**Fig. 9.** The transmission and reflection of light in the PMT measurement with (a) crystal perfectly matched to PMT window and (b) Thin optical grease layer coupled crystal to PMT window.

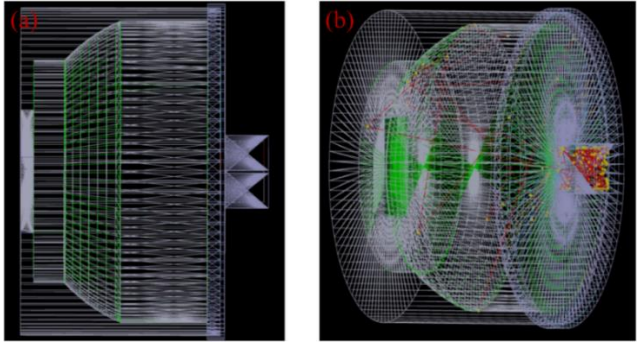

**Fig. 10.** The PMT reflectivity simulation using GEANT4 Optical photon: (a) geometry construction, and (b) passage of scintillation photons.

The intricacies of transmission and reflection of scintillation light are shown in Fig. 9. The scenario involving four layers has been explored in [40,41]. In Fig. 9(a), the crystal, PMT window, photocathode, and PMT vacuum possess refractive indices of $n_1$, $n_2$, $n_{ph}(1+ik_{ph})$, and $n_4$, respectively. The reflectance and transmittance are given by:

$$T = \frac{1}{2}\sum_{TM,TE}\frac{T_{12}T_3}{1-R_{12}R_3} \tag{7}$$

$$R = \frac{1}{2}\sum_{TM,TE}R_{12}+\frac{T_{12}^2R_3}{1-R_{12}R_3} \tag{8}$$

We extended the study to the scenario of five layers, the crystal, optical grease, PMT window, photocathode, and PMT vacuum will have a refractive index of $n_1$, $n_2$, $n_3$, $n_{ph}(1+ik_{ph})$, and $n_5$, respectively. The reflectance and transmittance are given by:

$$\begin{aligned}T &= \frac{1}{2}\sum_{TM,TE}\frac{T_{12}T_{23}T_4}{1-R_{12}R_{23}}\\ &+ \frac{T_{12}T_{23}T_4R_4}{1-R_{12}R_{23}}\frac{R_{23}+R_{12}T_{23}^2-R_{12}R_{23}^2}{(1-R_{12}R_{23})(1-R_{23}R_4)-R_{12}T_{23}^2R_4}\end{aligned} \tag{9}$$

$$\begin{aligned}R &= \frac{1}{2}\sum_{TM,TE}R_{12}+\frac{T_{12}^2R_{23}}{1-R_{12}R_{23}}\\ &+ \frac{T_{12}^2T_{23}^2R_4}{1-R_{12}R_{23}}\frac{1}{(1-R_{12}T_{23})(1-R_{23}R_4)-R_{12}T_{23}^2R_4}\end{aligned} \tag{10}$$

In the context of PMT, the scintillation photons can be absorbed, reflected, or transmitted in a photocathode. The transmitted photons can be reflected by reflective coating inside the PMT body to return to the photocathode, where absorption, reflection, or transmission can occur again. In the case photons inside PMT come back to the photocathode, the formula for the four layers is:

$$T = \frac{1}{2}\sum_{TM,TE}\frac{T_{12}T_3}{1-R_{12}R_3} \tag{11}$$

$$R = \frac{1}{2}\sum_{TM,TE}R_3+\frac{R_{12}T_3^2}{1-R_{12}R_3} \tag{12}$$

While for the case of five layers in Fig. 9(b), the formula is:

$$\begin{aligned}T &= \frac{1}{2}\sum_{TM,TE}\frac{T_{12}T_{23}T_4}{1-R_{23}R_4}\left[\frac{1-R_{12}R_{23}}{1-R_{12}R_{23}}\right.\\ &\left.+ \frac{R_{23}R_4(1-R_{12}R_{23})+R_{12}T_{23}^2R_4}{(1-R_{12}R_{23})-R_{12}T_{23}^2R_4}\right]\end{aligned} \tag{13}$$

$$\begin{aligned}R &= \frac{1}{2}\sum_{TM,TE}R_{45}+\frac{R_{23}T_4^2}{1-R_{23}R_4}\\ &+ \frac{R_{12}T_{23}^2T_4^2}{1-R_{23}R_4}\frac{1}{(1-T_{23}R_4)(1-R_{12}R_4)-R_{12}T_{23}^2R_4}\end{aligned} \tag{14}$$

T and R can be derived based on the surface transmission and reflection in Ref. [42]. It should be noted that total reflectance and transmittance must consider the transverse electric and magnetic waves. The formula can be fit for the case of small crystal – big PMT window or crystal with the same size as big PMT, which is fitted to our experiment with 3" PMT. The calculation of reflectance and transmittance are shown in Fig. 11.

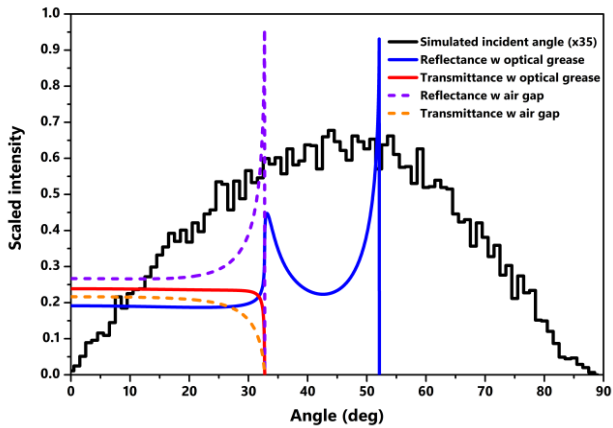

**Fig. 11.** Calculated reflectance and transmittance with optical grease and air gap. The simulated incident angle is from GEANT4.

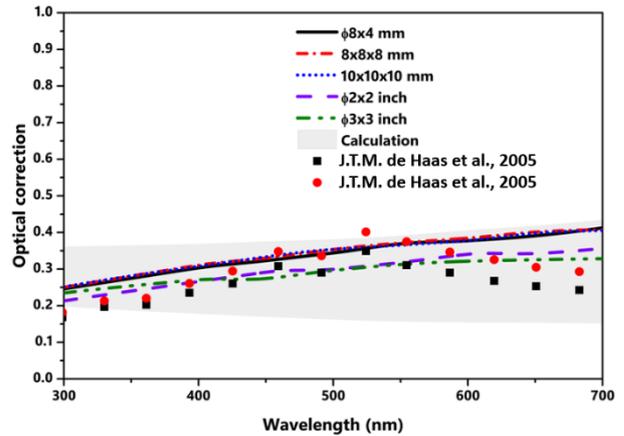

**Fig. 12.** The PMT optical correction factor from simulation and calculation.

Through calculations, we can determine the wavelength dependencies of reflectance and transmittance. However, a simulation can yield more comprehensive results, particularly concerning the setup geometry for a specific case. This study utilized the GEANT4 – Monte Carlo simulation toolkit [43]. The packages of the optical photon and scintillation have been applied. GEANT4 allows for generating the scintillation light inside crystal material under radiation interaction and tracking complete information of a single optical photon. By monitoring all generated scintillation photons concerning the track position and momentum, we can derive the optical properties of the photocathode in PMT. This makes it possible to study cases that are challenging for experimental measurement. In this work, we simulated NaI:Tl crystals of varying sizes a 3" bialkali PMT, complete with a Teflon cover and optical grease as the experiment. The construction of the geometry is illustrated in Fig. 10. The PMT was constructed with a window, a photocathode, an internal reflected area, and a dynode window excluding the geometry of the dynode system. The crystal was also constructed with an adjustable size and a Teflon cover. As shown in Fig. 12, the simulation closely aligns with the experiment. Based on the simulation, the angle distribution at the crystal-optical grease surface and the angle of the PMT vacuum-photocathode surface can be applied to obtain the calculated limited range for the PMT optical problem. In calculation and simulation, the thickness and the complex refractive index of the photocathode are essential, which were studied in [41,42].

The results from the simulation and calculation for NaI:Tl crystals are shown in Fig. 12, alongside the experimental reference data from [39]. These authors measured the PMT optical correction using the integrating sphere with the approach of diffuse reflectivity. The data from the reference falls within the calculation limit of 320 nm to 700 nm. Our simulations, as mentioned, accounted for the Teflon cover, crystal size with optical grease, and the scintillation light generation from radiation. There are discrepancies between simulation and reference data, which are not unexpected given the different methodological approaches. The reference data represents the diffuse reflectance measured using an integrating sphere with a 2" bialkali PMT with a quartz window from the light beam. It is suggested that optical grease should not be used in the measurement to apply the measured optical correction. This would not pose a problem if the crystal surface were perfectly attached to the PMT window. However, total reflection should be considered if there are air gaps. The calculations for the scenario involving optical grease and the rough case where an air layer exists between the crystal and PMT window are shown in Fig. 11. As can be observed from Fig. 11, optical grease allows more photons to reach the photocathode. However, this effect could be less pronounced if the air gap area is smaller. The total reflected photons can also reach the photocathode after numerous reflections made by Teflon. However, this occurs at the cost of potential photon absorption in the scintillator, window, and optical grease. It is important to note that the study on self-absorption in NaI:Tl crystals is not yet completed, while the absorption in window and optical grease can generally be disregarded.

## 5 Absolute light yield methodology

The absolute light yield is ascertained through the emitted photoelectrons, which can be measured by comparing the full peak photo-absorption ($Q_\gamma$) using γ-rays with energy $E_\gamma$ and the single photoelectron peak ($Q_{spe}$) [19,39]. As previously mentioned, the effective quantum efficiency must be corrected for the PMT optic. The absolute light yield is determined by the following equation:

$$Y_{ph} = \frac{Q_\gamma}{Q_{spe} E_\gamma} \frac{F_{abs} \int I(\lambda) d\lambda}{\int \frac{\epsilon_0(\lambda) I(\lambda)}{1 - F_{opt}(\lambda)} d\lambda} \qquad (15)$$

In this equation, $\epsilon_0$, $I$, $F_{opt}$, and $F_{abs}$ correspond to the manufacturer's quantum efficiency, X-ray emission intensity, optical correction, and absorption correction, respectively, all of which depend on the visible photon wavelength ($\lambda$). The X-ray emissions were measured with the same conditions following the scintillation characterization. Typical X-ray emissions are shown in Fig. 3 alongside the PMT manufacturer's quantum efficiency. In this study, the corrections are derived from the GEANT4 simulation. The reflection correction of the Teflon layer is a factor of 0.98 [44].

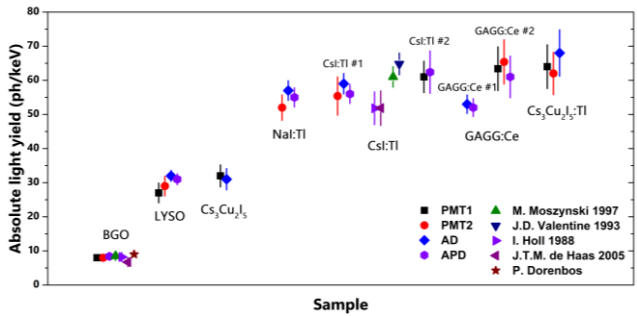

**Fig. 13.** The absolute light yields for several crystals.

**Table 3** The measured absolute light yield of different scintillators using PMT and PD/APD.

| Scintillation crystals | Absolute light yield (ph/keV) | | | |
|---|---|---|---|---|
| | R6233-100 PMT | R12669SEL PMT | PD | ADP |
| BGO | 8.0±0.6 | 8.0±0.3 | - | 8.4±0.4 |
| LYSO | 27.0±2.9 | 29.0±2.9 | 32.0±1.6 | 31.0±1.6 |
| GAGG:Ce #1 | 63.4±6.4 | 65.4±6.5 | - | 61.0±6.1 |
| GAGG:Ce #2 | - | - | 53.0±2.7 | 52.0±2.6 |
| CsI:Tl #2 | 61.0±4.6 | - | - | 62.4±6.2 |
| CsI:Tl #2 | - | 55.4±5.6 | 59.0±3.0 | 56.0±2.8 |
| Cs₃Cu₂I₅:Tl | 64.0±6.4 | 62.0±6.2 | 68.0±6.8 | - |
| Cs₃Cu₂I₅ | 32.0±3.2 | - | 31.0±3.1 | - |
| Encapsulated ASWSII-2 NaI:Tl | - | 52.0±3.7 | 57.0±2.9 | 55.0±2.8 |

Evaluating systematic error is crucial for assessing the reliability of the SPE method. Therefore, this work compared the absolute light yield measured using the PMTs and photodiodes (PD and APD). For the PMT method, we used another PMT, Hamamatsu 3" R6233-100 PMT, for comparison with Hamamatsu 3" R12669SEL PMT. In the case of the photodiode method, the absolute light yield was measured using a method that can be seen in Ref. [41,45]. Their results are listed in Table 3 and plotted in Fig. 13 compared to other data [17,39,46-48]. From Table 3, we can infer that a systematic error of 10% can be assigned among measurements using different methods and equipment. This systematic error will be applied to the results of NaI:Tl crystals. Furthermore, when using the Hamamatsu 3" R12669SEL PMT, the fluctuation in absolute light yield is less than 3% across a voltage range from 900 V to 1200 V. This observation underscores the stability of our algorithm in determining the single photoelectron response for measuring absolute light yield measurement, especially at low gain PMT operation as previously mentioned.

## 6 Absolute light yields of NaI:Tl crystals

**Table 4** The measured absolute light yield of NaI:Tl crystals.

| Scintillation crystals | Dimension (mm³) | Absolute light yield (ph/keV) |
|---|---|---|
| KNU+COSINE R&D | φ8x4 – 10x10x10 | 31.0±3.1 - 58.5±5.9 |
| Bare ASWSII-2 | 8x8x8 | 57.40±5.7 |
| Encapsulated ASWSII-2 | 8x8x8 | 52.0±5.2 |
| Bare CUP KY01-2007 | 8x8x8 | 50.0±5.0 – 59.4±5.9 |
| Bare CUP 036 | 8x8x8 | 44.0±4.4 |
| Bare Alpha Spectra 1 | 10x10x10 | 54.0±5.4 |
| Bare Alpha Spectra 2-1 | 10x10x10 | 55.0±5.5 |
| Bare Alpha Spectra 2-2 | 10x10x10 | 54.0±5.4 |
| Bare Alpha Spectra 2-3 | 10x10x10 | 58.0±5.8 |
| Bare Alpha Spectra 2-4 | 10x10x10 | 51.3±5.1 |
| Bare EPIC | φ50.8x50.8 | 49.5±5.0 |
| Encapsulated EPIC | φ50.8x50.8 | 32.4±3.2 |
| Bare BICRON sample | φ76.2x76.2 | 25.0±2.5 |
| Encapsulated Alpha Spectra 1 | φ76.2x76.2 | 42.0±4.2 |
| Encapsulated Alpha Spectra 2 | φ76.2x76.2 | 48.0±4.8 |

As measured, the absolute light yields of various NaI:Tl samples are detailed in Table 4 and compared with other data in Fig. 14. The values of absolute light yield vary a wide range from 20 ph/keV to 60 ph/keV. This variation underscores the challenges associated with the growth of NaI:Tl crystals. In this study, the absolute light yield of small crystals ranges from 50 ph/keV to 60 ph/keV, except for the CUP 036 crystal. Almost all measured and referred absolute light yields surpass the 38 ph/keV reported from NaI:Tl crystal Saint-Gobain [26]. In the case of BICRON crystal, it was extracted from the compact detector. The crystal lacks cracks, but it exhibits some irreparable damage, such as several small holes from the bottom and extending deep into the center, the cause of which remains unknown. The bare EPIC crystal was also extracted from the Aluminum encapsulated scintillator due to the degraded surface inside the case. After the polishing, this crystal shows 49.5±5.0 ph/keV. As observed in Fig. 14, the COSINE-100 and Alpha spectra samples boast an excellent light yield, peaking at approximately 59.4±5.9 ph/keV. Based on the NaI:Tl light output measurement at -35 °C, light yield can be improved by 4.7%, suggesting that these crystals might achieve a light yield of 62.0±6.2 ph/keV [49].

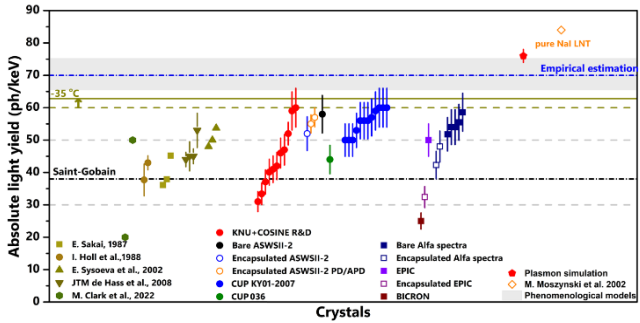

**Fig. 14.** Comparison of the NaI:Tl absolute light yields.

The highest measured value is close to the empirical estimation of the maximum light yield of NaI:Tl [21]. Furthermore, the Robbins and Callaway phenomenological models show that the maximum electron-hole pairs (eh) generated in NaI:Tl ranges from 65.4 eh/keV to 75.3 eh/keV [50]. The plasmon simulation using GEAN4 for NaI at room temperature shows the maximum of eh pairs to be approximately 76 eh/keV [51]. The relationship between the light yield ($Y_{ph}$) and electron-hole pair yield ($Y_{eh}$) can be derived as following:

$$Y_{ph} = Y_{eh}SQ \qquad (16)$$

In which, the S and Q are the efficiencies of transfer and luminescence processes, respectively. In general, the Q value could be measured in optical spectroscopy of solids [52]. In the case of NaI:Tl it is considered as unity [52]. While the S value is obtained by comparing the values of $N_{ph}$ and $N_{eh}Q$. Assuming the maximum transfer efficiency, these values can be utilized to estimate the maximum light yield of NaI:Tl crystal. As shown in Fig. 14, the measured absolute light yields excited by 661 keV γ rays are near this range, especially for estimation at -35 °C based on Ref. [49]. It could be supposed that the 70±7 ph/keV could the limitation of NaI:Tl crystal. Therefore, the transfer efficiency of NaI:Tl crystal could reach 0.85±0.09 at room temperature and 0.89±0.09 at -35 °C which are higher than the values of 0.5 [50] and 0.59 [52] based on the NaI:Tl light yield of 38 ph/keV. If we consider the non-proportionality of NaI:Tl crystal which is higher approximately 12% at 59.5 keV γ-ray excitation [53], we will have light yield values of 67±7 ph/keV at room temperature and 70±7 ph/keV at -35 °C. It means that our results agree with current electron-hole pair models with transfer efficiency values close to unity.

Comparing the measurement of Ref. [54] at liquid nitrogen temperature for three pure NaI crystals excited by 661 keV γ rays, their light yields are 69±7 ph/keV, 84±9 ph/keV, and 62±7 ph/keV for sample 1, 2, and 3, respectively. It should be noted that correction factor of 0.58±0.03 was introduced due to the light losses of their setup [54]. Sample 1 is the almost clean pure NaI crystal which light yield is close to the electron-hole pair predictions with approximately unity efficiencies. For the case of sample 2, there was an unknown 380 nm emission. While sample 3 had thallium contamination. Our highest results are close to the light yield values of sample 1 and 3 under uncertainty. It suggests high efficiency of scintillation processes. However, the light yield of 84±9 ph/keV exhibited by the pure NaI crystal at liquid nitrogen temperature exceeds the theoretical light yield range. This discrepancy raises the question: Is it possible for the light yield of NaI:Tl light yield up to a range of 75-85 ph/keV under 661 keV γ-ray excitation?

In the case of the CUP KY01-2007 samples from the COSINE-100 experiment, the large slide crystal was segmented into several pieces of similar size. The distribution of light yield can be seen in Fig. 15(a). The X-ray emission spectra, as shown in Fig. 15(b), implies consistency in emission at different positions in big crystals. However, the absolute light yields vary up to 20% at different positions, as shown in Fig. 15(a). The light yield has a high value at the top, decreasing from top to bottom and center to border. It could reduce the absolute light yield of a larger size to around 55.0 ph/keV. This effect could be attributed to the distribution of thallium during the crystal growth processes. Besides, the light yield in KNU+COSINE crystals vary from 31.0±3.1 ph/keV to 58.5±5.9 ph/keV at 661 keV γ-ray excitation. The observation in CUP KY01-2007 and KNU+COSINE samples reaffirm the strong effect of growth quality and technique on the absolute light yield of NaI:Tl crystal. In addition, the absolute light yield measurement with accuracy is necessary for NaI:Tl crystals for calibration or reference purposes.

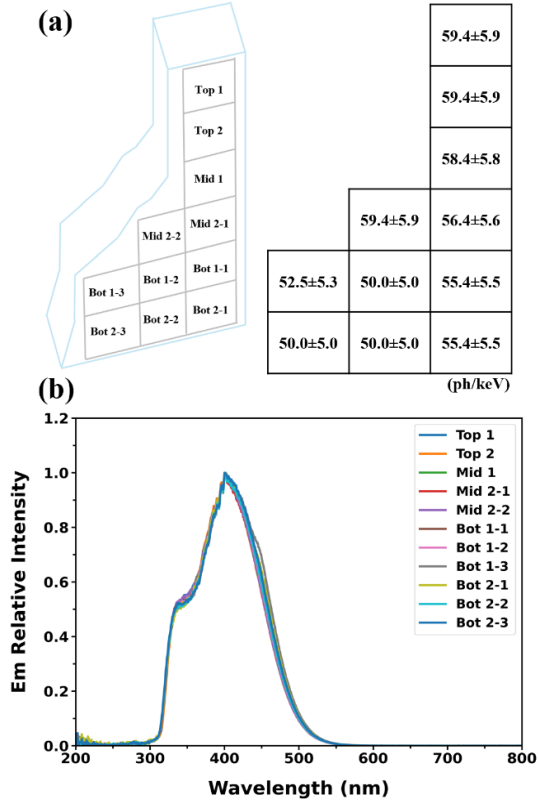

**(a)**

| | | | 59.4±5.9 |
|---|---|---|---|
| Top 1 | | | 59.4±5.9 |
| Top 2 | | | 58.4±5.8 |
| Mid 1 | | | |
| Mid 2-1 | Mid 2-2 | 59.4±5.9 | 56.4±5.6 |
| Bot 1-1 | | | |
| Bot 1-2 | 52.5±5.3 | 50.0±5.0 | 55.4±5.5 |
| Bot 1-3 | | | |
| Bot 2-1 | | | |
| Bot 2-2 | 50.0±5.0 | 50.0±5.0 | 55.4±5.5 |
| Bot 2-3 | | | (ph/keV) |

**(b)**

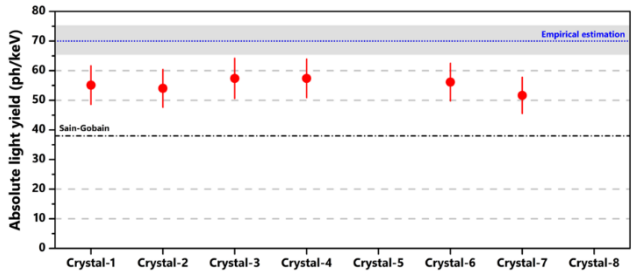

**Fig. 15.** The information of CUP KY01-2007 samples: (a) positions and absolute light yields of samples (ph/keV) of small samples, (b) The X-ray emissions of these samples.

**Fig. 16.** The light yield evaluation of 8 crystals used in the COSINE-100 experiment.

The 3" encapsulated crystal from Alpha spectra exhibits a commendable light yield. Assuming they have been grown under the same conditions as the small bare Alpha spectra samples, the light loss is from 7% to 28%. The loss is 35% for EPIC samples with the same assumption. The measurement of light loss after encapsulation for the ASWSII-2 sample is 10%. This crystal can be utilized to evaluate the COSINE crystals, as reported in Ref. [8], which is in the photoelectron yield ($Y_{pe}$ pe/keV). In Ref. [8], crystal-3 and crystal-4 have the same material as the ASWSII-2 sample. The same PMT was used in Ref. [8] and this work. We could estimate the light loss (based on $Y_{pe}$ pe/keV) based on a bare ASWSII-2 sample, which has 24.3±1.7 pe/keV at 661 keV or approximately 27.2±1.9 pe/keV at 59.5 keV with approximately 12% of NaI:Tl light yield non-proportionality between 59.5 keV and 661 keV [53]. Meanwhile, crystal-3 (ASWSII) has 15.5±1.6 pe/keV at 59.5 keV, and crystal-4 (ASWSII) has 14.9±1.5 pe/keV at 59.5 keV. The absolute light yields of crystal-3 and crystal-4 are 48.0±4.8 ph/keV and 46.0±4.6 ph/keV, respectively. The light loss is estimated at around 40-46% compared to the small, bare ASWSII-2 sample. By assuming the main light losses of encapsulated crystal is due the encapsulation technique, the evaluation of 6 out of 8 current crystals in the COSINE-100 experiment can be carried out. Their results are shown in Fig. 16. Crystal-1, crystal-2, crystal-3, crystal-4, crystal-6 and crystal-7 [8] have a high light yield from 51 ph/keV to 57 ph/keV. The other two crystals including crystal-5 and crystal-8 [8] are out of our estimation condition due to different encapsulation techniques. This suggests that while the COSINE-100 experiment employs excellent crystals, only more than half of the potential is exploited. Although the study of bulk NaI:Tl self-absorption is not yet complete, achieving a light output improvement of 40-46% is challenging. Nevertheless, investigation in R&D for encapsulation techniques to enhance light yield could prove beneficial.

# 7 Conclusion

This work used the single photoelectron method to measure the absolute light yield of various NaI:Tl crystals. The theoretical estimation and detailed geometry simulation have been performed to study the optics of PMT. Results show that a factor of 27% needs to be corrected for the manufacture quantum efficiency of PMT. The photon absorption in large-size crystals has been corrected. The absolute light yield will be overrated without correction of quantum efficiency related to the PMT optic. Besides, due to the wavelength dependence of correction, the comparison methods for light yield measurement also considered the PMT optic if the reference and measured crystal do not match the wavelength range. These corrections have been used in our measurements. The different PMTs and methods (Si-photodiode and Si-Avalanche photodiode methods) have been performed with different types of crystals. The consistency among these results confirms the reliability of the absolute light yield measurement using PMT for NaI:Tl crystals with 10% systematic error. The SPE method has been applied to measure the absolute light yield of 35 NaI:Tl samples of various sizes. Among them, some crystals have been used or developed for the COSINE-100 experiment. The absolute light yields vary almost from 50.0 ph/keV to 60.0 ph/keV. Excellent values have been obtained at 59.4 ph/keV, close to the theoretical maximum light yield range, which implies high transfer efficiency. This result is approximately 50% higher than the typically quoted value. By comparing bare and encapsulated crystals, the light loss of encapsulation could be a severe problem in the big-size scintillator, which can reduce more than 20% of light yield. Based on the rough estimation, the absolute light yields of the six current crystals in the COSINE-100 experiment are high light yields. Our estimations suggest improving the encapsulation technique could enhance their light output (pe/keV). Besides, the NaI:Tl R&D in the COSINE-100 experiment shows the reachable value of 59.4 ph/MeV for a small sample.

## Acknowledgements

This work is supported by Institute for Basic Science (IBS) and the National Research Foundation of Korea funded by the Ministry of Science and ICT, Korea (Under Grant: IBS-R016-D1, and 2018R1A6A1A06024970). We would like to thank the COSINE-100 experiment for the many NaI:Tl samples and equipment. We would like to thank Alpha Spectra for the several NaI:Tl samples. We would like to thank the KIMS experiment for the CsI:Tl samples.

## References

1  L. Roszkowski, E. M. Sessolo, and S. Trojanowski, "WIMP dark matter candidates and searches—current status and future prospects," Reports on Progress in Physics, 81(6) (2018) 066201, 2018. https://doi.org/10.1088/1361-6633/aab913

2  XENON Collaboration, "The XENON1T dark matter experiment," The European Physical Journal C, 77 (2017) 881. https://doi.org/10.1140/epjc/s10052-017-5326-3

3  J. Amaré et al., "Dark Matter Searches Using NaI(Tl) at the Canfranc Underground Laboratory: Past, Present and Future," Universe, 8(2) (2022) 75. https://doi.org/10.3390/universe8020075

4  Y. S. Yoon et al., "Search for solar axions with CsI(Tl) crystal detectors," Journal of High Energy Physics, 2016(6) (2016) 11. https://doi.org/10.1007/JHEP06(2016)011

5  E. Barbosa de Souza et al., "First search for a dark matter annual modulation signal with NaI(Tl) in the Southern Hemisphere by DM-Ice17," Physical Review D, 95(3) (2017) 032006. https://doi.org/10.1103/PhysRevD.95.032006

6  B. Ahmed et al., "The NAIAD experiment for WIMP searches at Boulby mine and recent results", Astroparticle Physics, 19(6) (2003) 691-702. https://doi.org/10.1016/S0927-6505(03)00115-4

7  COSINE-100 Collaboration, "Three-year annual modulation search with COSINE-100," Physical Review D, 106(5) (2022) 052005. https://doi.org/10.1103/PhysRevD.106.052005

8  COSINE-100 Collaboration, "Initial performance of the COSINE-100 experiment," The European Physical Journal C, 78(2) (2018) 107. https://doi.org/10.1140/epjc/s10052-018-5590-x

9  J. Amaré et al., "Annual modulation results from three-year exposure of ANAIS-112," Physical Review D, 103(10) (2021) 102005. https://doi.org/10.1103/PhysRevD.103.102005

10  COSINE-100 Collaboration, "Strong constraints from COSINE-100 on the DAMA dark matter results using the same sodium iodide target," Science Advances, 7(46) (2021). https://doi.org/10.1126/sciadv.abk2699

11  K. Fushimi et al., "Dark matter search with high purity NaI(Tl) scintillator", arXiv (2021). https://doi.org/10.48550/arXiv.2106.15235

12  Y.J. Ko and H.S. Lee, "Sensitivities for coherent elastic scattering of solar and supernova neutrinos with future NaI(Tl) dark matter search detectors of COSINE-200/1T", Astroparticle Physics, 153 (2023) 102890. https://doi.org/10.1016/j.astropartphys.2023.102890

13  M. Antonello et al., "The SABRE project and the SABRE Proof-of-Principle," The European Physical Journal C, 79 (4) (2019) 363. https://doi.org/10.1140/epjc/s10052-019-6860-y

14  K. Fushimi et al., "Dark matter search project PICO-LON," J Phys Conf Ser, 718 (2016) 042022. https://doi.org/10.1088/1742-6596/718/4/042022

15  G. Angloher et al., "COSINUS: Cryogenic Calorimeters for the Direct Dark Matter Search with NaI Crystals," J Low Temp Phys, 200(5–6) (2020) 428–436. https://doi.org/10.1007/s10909-020-02464-9

16  E. Sakai, "Recent measurements on scintillator-photodetector systems," IEEE Transactions on Nuclear Science 34 (01) (1987) 418-422. https://doi.org/10.1109/TNS.1987.4337375

17  I. Holl, E. Lorenz, and G. Mageras, "A measurement of the light yield of common inorganic scintillators," IEEE Transactions on Nuclear Science 35 (01) (1988) 105-109. https://doi.org/10.1109/23.12684

18  E. Sysoeva, V Tarasov, and O. Zelenskaya, "Comparison of the methods for determination of scintillation light yield", Nuclear Instruments and Methods in Physics Research Section A: Accelerators, Spectrometers, Detectors, and Associated Equipment 486 (1-2) (2002) 67-73. https://doi.org/10.1016/S0168-9002(02)00676-9

19  JTM De Haas, and P. Dorenbos, "Advances in yield calibration of scintillators", IEEE Transactions on Nuclear Science 55 (3) (2008) 1086-1092. https://doi.org/10.1109/TNS.2008.922819

20  M. Clark, F. Froborg, P.C.F. Di Stefano and F. Calaprice, "Investigation of the cryogenic scintillation of pure and doped sodium-iodine," Journal of Instrumentation 17 (05) (2022) P05018. https://doi.org/10.1088/1748-0221/17/05/P05018

21 K. Kramer, P. Dorenbos, H.U. Gudel, and C.W.E. Van Eijk, "Development and characterization of highly efficient new cerium doped rare earth halide scintillator materials," Journal of Materials Chemistry 16 (27) (2006) 2773-2780. https://doi.org/10.1039/B602762H

22 K. Shin et al., "A facility for mass production of ultra-pure NaI powder for the COSINE-200 experiment," Journal of Instrumentation 15 (2020) C07031. https://doi.org/10.1088/1748-0221/15/07/C07031

23 B.J. Park et al., "Development of ultra-pure NaI(Tl) detectors for the COSINE-200 experiment," The European Physical Journal C 80 (2020) 814. https://doi.org/10.1140/epjc/s10052-020-8386-8

24 Epic Crystal Co., available at https://www.epic-crystal.com/

25 Alpha Spectra Inc., available at https://alphaspectra.com/

26 Sain-Goban, USA, available at https://www.crystals.saint-gobain.com/.

27 ELJEN technology, available at https://eljentechnology.com/

28 H.S. Lee et al., "Development of low-background CsI(Tl) crystals for WIMP search," Nuclear Instruments and Methods in Physics Research Section A: Accelerators, Spectrometers, Detectors, and Associated Equipment 571 (3) (2007) 644-650. https://doi.org/10.1016/j.nima.2006.10.325

29 Notice Korea, available at http://www.noticekorea.com/

30 R. Brun and F. Rademakers, "ROOT - An Object-Oriented Data Analysis Framework," Nuclear Instruments and Methods in Physics Research Section A: Accelerators, Spectrometers, Detectors, and Associated Equipment 389 (1997) 81–86. https://doi.org/10.1016/S0168-9002(97)00048-X

31 ORTEC, available at https://www.ortec-online.com/

32 J. Pivarski, Uproot: Minimalist root i/o in pure Python and numpy (2017), available at https://github.com/scikit-hep/uproot5

33 Ocean Insight, available at https://www.oceaninsight.com/

34 Teledyne Lecroy, available at https://www.teledynelecroy.com/

35 Hamamatsu, available at https://www.hamamatsu.com/us/en.html

36 R. Saldanha et al., "Model independent approach to the single photoelectron calibration of photomultiplier tubes," Nuclear Instruments and methods in Physics Research A, 863 (2017) 35-46. https://doi.org/10.1016/j.nima.2017.02.086

37 R. Dossi, A. Ianni, G. Ranucci, and O.Ju. Smirnov, "Methods for precise photoelectron counting with photomultipliers," Nuclear Instruments and Methods in Physics Research Section A: Accelerators, Spectrometers, Detectors, and Associated Equipment 451 (2000) 623-637. https://doi.org/10.1016/S0168-9002(00)00337-5

38 P. Besson, Ph. Bourgeois, P. Garganne, J.P. Robert, L. Giry, and Y. Vitel, "Measurement of photomultiplier quantum efficiency," Nuclear Instruments and Methods in Physics Research Section A: Accelerators, Spectrometers, Detectors, and Associated Equipment 334 (1994) 435-437. https://doi.org/10.1016/0168-9002(94)90095-7

39 J.T.M. De Haas, P. Dorenbos, and C.W.E. Van Eijk, "Measuring the absolute light yield of scintillators," Nuclear Instruments and Methods in Physics Research Section A: Accelerators,

40 D. Motta, and S. Schonert, "Optical properties of bialkali photocathodes," Nuclear Instruments and Methods in Physics Research Section A: Accelerators, Spectrometers, Detectors, and Associated Equipment 539 (2005) 217-235. https://doi.org/10.1016/j.nima.2004.10.009

41 M.E. Moorhead, and N.W. Tanner, "Optical properties of an EMI $K_2CsSb$ bialkali photocathode," Nuclear Instruments and Methods in Physics Research Section A: Accelerators, Spectrometers, Detectors, and Associated Equipment 378 (1996) 162-170. https://doi.org/10.1016/0168-9002(96)00447-0

42 M. Born, E. Wolf, Principle of Optics, Pergamon Press, Oxford, 1964.

43 S. Agostinnel et al., "GEANT4 – a simulation toolkit," Nuclear Instruments and Methods in Physics Research Section A: Accelerators, Spectrometers, Detectors, and Associated Equipment 506 (3) (2003) 250-303. https://doi.org/10.1016/S0168-9002(03)01368-8

44 V.R. Weidner, and J.J. Hsia, "Reflection properties of pressed polytetrafluoroethylene powder," Journal of the Optical Society of America 71-7 (1981) 856-861. https://doi.org/10.1364/JOSA.71.000856

45 J. Jegal, H.W. Park, H. Park, and H.J. Kim, "Development of the calibration system and characterization of the Avalanche photodiode for scintillation detection," IEEE Transactions on Nuclear Science 68-6 (2021) 1304-1308. https://doi.org/10.1109/TNS.2021.3070040

46 J.D. Valentine et al. "Temperature dependence of CsI(Tl) absolute scintillation yield," IEEE Transactions on Nuclear Science, 40(4) (1993) 1267-1274. https://doi.org/10.1109/TNS.1993.8526779

47 M. Moszynski et al., "Absolute light output of scintillators," IEEE Transactions on Nuclear Science, 44(3) (1997) 1052-1061. https://doi.org/10.1109/23.603803

48 P. Dorenbos et al., "non-proportionality in the scintillation response and the energy resolution obtainable with scintillation crystals," IEEE Transactions on Nuclear Science, 42(6) (1995) 2190-2202. https://doi.org/10.1109/23.489415

49 S.H. Lee, G.S. Kim, H.J. Kim, K.W. Kim, J.Y. Lee, and H.S. Lee, "Study on NaI(Tl) crystal at -35 ℃ for dark matter detection," Astroparticle Physics 141 (2022) 10279. https://doi.org/10.1016/j.astropartphys.2022.102709

50 R.H. Bartram, and A. Lempicki, "Efficiency of electron-hole pair production in scintillators," Journal of Luminescence 68 (996) 225-240. https://doi.org/10.1016/0022-2313(96)00026-9

51 N.T. Luan et al., "Simulation of Electron-Hole Pairs Created by γ-Rays Interaction in Scintillation Crystals Using Geant4 Toolkit", IEEE Transactions on Nuclear Science, 70(7) (2023) 1312-1317. https://doi.org/10.1109/TNS.2023.3271328

52 A. Lempicki, A.J. Wojtowicz, and E. Berman, "Fundamental limits of scintillator performance," Nuclear Instruments and Methods in Physics Research Section A: Accelerators, Spectrometers, Detectors, and Associated Equipment 333 (1993) 304-311. https://doi.org/10.1016/0168-9002(93)91170-R

53  L. Swiderski, M. Moszynski, W. Czarnacki, A. Syntfeld-Kazuch, and M. Gierlik, "Non-proportionality and Energy Resolution of NaI(Tl) at Wide Temperature Range (-40°C to +23°C)," in 2006 IEEE Nuclear Science Symposium Conference Record, (2006) 1122–1128. https://doi.org/10.1109/NSSMIC.2006.356043

54  M. Moszynski et al., "Study of pure NaI at room and liquid nitrogen temperatures," in 2002 IEEE Nuclear Science Symposium Conference Record, (2002) 346–351. https://doi.org/10.1109/NSSMIC.2002.1239330